\documentclass[aps,prl,reprint,onecolumn,showpacs,superscriptaddress,longbibliography,10pt]{revtex4-1}
\usepackage[english]{babel}
\usepackage{bm}
\usepackage{amsfonts}

\usepackage{graphicx}         
\usepackage{amsmath}         
\usepackage{amssymb}          

\usepackage{soul}      

\usepackage{braket}

\usepackage[usenames,dvipsnames]{color}

\usepackage{siunitx}

\usepackage{subscript}

\begin{document}

\title{Roton excitations and the fluid-solid phase transition
in superfluid 2D Yukawa Bosons}

\author{S. Molinelli}
\affiliation{Dipartimento di Fisica, Università degli Studi di Milano, via Celoria 16, 20133 Milano, Italy}
\author{D.E. Galli}
\affiliation{Dipartimento di Fisica, Università degli Studi di Milano, via Celoria 16, 20133 Milano, Italy}
\author{L. Reatto}
\affiliation{via Bazzini 20, 20131 Milano, Italy}
\author{M. Motta}
\affiliation{Department of Physics, The College of William and Mary, Williamsburg VA, 23187, USA}

\begin{abstract}
We compute several ground–state properties and the dynamical
structure factor of a $0$--temperature system of Bosons interacting with the
2D screened Coulomb (2D-SC) potential. We resort to the exact shadow path--integral
ground--state (SPIGS) quantum Monte Carlo method to compute the imaginary--time 
correlation function of the model, and to the genetic algorithm via falsification of 
theories (GIFT) to retrieve the dynamical structure factor. We provide a detailed comparison
of ground–state properties and collective excitations of 2D-SC and $^4$He atoms. 
The roton energy of the 2D-SC system is an increasing function of density, and not 
a decreasing one as in $^4$He. This result is in contrast with 
the view that the roton is the soft mode of the fluid-solid transition. We uncover a 
remarkable quasi-universality of backflow and of other properties when expressed in 
terms of the amount of short range order as quantified by the height of the first peak
of the static structure factor.

\end{abstract}


\maketitle

\section{Introduction}
\label{intro}

While $^4$He has long been the archetypal Bosonic quantum fluid \cite{Pines1966,Fetter1971,Glyde1994,Ceperley1995}, 
the last few decades have witnessed a considerable increase in the number of relevant Bosonic fluids available for 
experiments \cite{Bloch2008,Cazalilla2011}. An important issue is determining the extent to which the properties of 
a Bose fluid are affected by the details of the inter--atomic interaction, $V(r)$. 
In the case of cold atoms at very low densities it is known \cite{Bloch2008} that under most conditions only one 
parameter of such interaction plays a role in determining the properties of the system, namely the scattering length 
at low energy. The situation is different for systems at large density, in which case $V(r)$ explicitly enters 
into the microscopic description of the system. 
A number of model interactions have been studied by many--body theories and simulations \cite{Ceperley1995,Cazalilla2011,DePalo2004,Rota2013}. 
The He--He interaction is characterized by a hard repulsive core at short distance and by a rather weak attraction 
at larger distances, as typified by the Lennard--Jones form.
In fact, it is well-established that most properties of such system are determined by the hard core, the attractive
tail playing a minor role \cite{Kalos1981,Rota2013,Rota2014}.

One--component charged Bose fluids differ from the previous case in two ways: the weak divergence of $V(r)$ at $r=0$, allowing particles to overlap, and the long range of the interaction.
The screened Coulomb interaction, also called Yukawa interaction, combines in an 
interesting way a soft short--range repulsion and a finite range.  
Also, the interest of this last system in two dimensions (2D)
relies in a deep and truly fascinating connection 
with the thermodynamics of Abrikosov vortices in Type-II superconductors~\cite{Nelson1989}. 
The $T=0$ K phase diagram of the 2D-Screned Coulomb (2D-SC) Bose system is rather interesting. 
In a hard core system like $^4$He there is a phase transition from a superfluid state 
at low density to a solid phase at high density. On the opposite, the Coulomb system 
has a solid phase at low density and  a fluid phase at high 
density \cite{Giuliani2005,DePalo2004}.
The 2D-SC Bose system has features of both those systems. 
When the mass of the particles is sufficiently small, the kinetic energy dominates over the 
potential energy and quantum fluctuations are enhanced, whence the system is fluid at all densities.
Below a critical threshold, the system is solid for intermediate values of the 
density and fluid at both very low and high density.
The 2D-SC Bose system at $T=0$ K has been studied by variational theory \cite{Xing1990} 
and by exact ground--state methods, diffusion Monte Carlo \cite{Magro1993} 
and shadow path integral ground state (SPIGS) simulations \cite{Rossi2011}, 
and by exact finite-temperature path integral Monte Carlo simulations \cite{Nordborg1997,Nordborg1998}. 
The main focus of those studies was the phase diagram of the system, and the 
calculation of ground--state equilibrium properties.
In previous work \cite{Nordborg1997,Nordborg1998} only a scant attention was given to dynamical properties.
In this work we present the first study of dynamical properties of the 2D-SC Bose 
system in the fluid phase using state--of--the--art quantum Monte Carlo and 
analytic continuation techniques.

The dynamical structure factor of a many-body system unveils considerable information about the collective excitations 
of the system as probed by external perturbations coupling to the density \cite{Pines1966,Fetter1971}.
For a $T=0$ K system, the dynamical structure factor, $S({\bf{q}},\omega)$, 
is defined as the Fourier transform of the intermediate scattering function $F({\bf{q}},t)$,
which is the correlation function of the Fourier transform of the density operator, 
$\rho_{\bf{q}} = \sum_i e^{-i {\bf{q}} \cdot {\bf{r}}_i}$,
\begin{equation}
F({\bf{q}},t) = \frac{1}{N}\braket{\Psi_0|\rho_{\bf{q}}(t) \rho_{-{\bf{q}}}|\Psi_0}
= \frac{1}{N} \braket{\Psi_0|e^{\frac{it}{\hbar} H}\rho_{\bf{q}}
e^{-\frac{it}{\hbar}H} \rho_{-{\bf{q}}}|\Psi_0}
\end{equation}
$N$ being the number of particles, $H$ the Hamiltonian of the system and 
$\Psi_0$ the ground state of $H$.
Therefore,
\begin{equation}
S({\bf{q}},\omega) = \int dt \, \frac{e^{i \omega t}}{2\pi} \, F({\bf{q}},t) \quad .
\end{equation}
Within the Born approximation, the dynamical structure factor is related to the 
differential scattering cross section 
of particle or electromagnetic radiation through the multiplication by a purely kinematic factor \cite{Pines1966,Fetter1971,Glyde1994}.
This important property makes the dynamical structure factor a quantity of paramount importance in the
experimental and theoretical investigation of many-body quantum systems, ranging from superfluid and solid
${}^4$He \cite{Boninsegni1998,Moroni1998,Vitali2010,Rossi2012,Fak2012,Arrigoni2013} to ultracold 
atomic gases \cite{Pitaevskii2003,Rota2013}.

For strongly--interacting systems, the quantitative study of the dynamical structure factor must typically be
supported by numerical calculations or simulations.
The first application of quantum Monte Carlo (QMC) methods to the calculation of the dynamical structure factor 
dates back to \cite{Silver1990,Jarrell1996,White1991} for lattice systems and \cite{Boninsegni1996} for homogeneous systems.
Further developments have made possible the calculation of $S({\bf{q}},\omega)$ in systems of ${}^4$He atoms 
\cite{Vitali2010,Arrigoni2013}, as well as in two--dimensional (2D) homogeneous electron gases and systems 
of ${}^3$He atoms \cite{Nava2013,Motta2014,Motta2015}, and systems of Bose hard spheres \cite{Rota2013,Rota2014}.

In the present work, we compute the dynamical structure factor of a zero--temperature 
system of 2D-SC Bosons, using the shadow path integral ground state method (SPIGS)
\cite{Galli2003,Galli2004} to compute the imaginary--time intermediate scattering 
function, $F({\bf{q}},\tau)$ (see equation \eqref{eq:pigs-1}), and a genetic algorithm, known as genetic inversion via falsification 
of theories (GIFT) \cite{Vitali2010,Rota2013}, to retrieve $S({\bf{q}},\omega)$ from $F({\bf{q}},\tau)$.

The structure of the paper is as follows. In Section \ref{sec:2}, the model of 2D-SC Bosons is introduced and the 
SPIGS method is briefly reviewed.
In Subection \ref{sec:2b}, the calculation of $F({\bf{q}},\tau)$ and its inversion by means of the GIFT algorithm are described.
Results for ground state properties are reported in Section \ref{sec:3}, and those for the dynamical structure factor 
in Section \ref{sec:4}.
Conclusions are drawn in the last Section \ref{sec:5}. 

\section{Model and methods}
\label{sec:2}

The system considered in the present work consists of $N$ identical spinless Bosons of mass $m$, strictly confined
on a plane and interacting through the 2D screened Coulomb potential
\begin{equation}
V(r) = \epsilon \, K_0\left( \frac{r}{\sigma} \right)
\quad ,
\end{equation}
where $K_0(r)$ is a modified Bessel function of the second kind.
For small $r$, $K_0$ diverges as $-\ln(r)$; for large $r$ it decays as $e^{-r}/\sqrt{r}$.
Throughout the present work, lengths are measured in units of the screening length $\sigma$ and energies in units of 
the interaction strength $\epsilon$, whence the Hamiltonian reads
\begin{equation}
H = - \left( \Lambda^* \right)^2 \sum_{i=1}^N \triangle_i + \sum_{i<j=1}^N K_0(r_{ij}) \quad .
\end{equation}
The model contains two dimensionless parameters: the DeBoer parameter $\Lambda^*$, that controls the relative 
importance of the kinetic and potential energy and is defined by
\begin{equation}
\left( \Lambda^* \right)^2 = \frac{\hbar^2}{2m\sigma^2 \epsilon} \quad ,
\end{equation}
and the reduced density $\rho^* = \frac{N \sigma^2}{L_x \times L_y}$. The system has been simulated in 
a 2D rectangular box, with sides $L_x$ and $L_y$ (aspect ratio $L_y/L_x \simeq 1.1$) 
and periodic boundary conditions are enforced to reduce finite--size effects.
Also, nearest neighbors images and potential cut--off at $L_x/2$ have been implemented.

\subsection{The PIGS method}
\label{sec:2a}

QMC methods have been applied to the investigation of ground--state properties of 
2D-SC Bosons by several authors including Xing and Tesanovic \cite{Xing1990}, 
Magro and Ceperley \cite{Magro1993}, 
Nordborg and Blatter \cite{Nordborg1997}, Rossi {\em{et al}} \cite{Rossi2011}, 
achieving a semi--quantitative characterization of the zero--temperature phase diagram 
and off--diagonal properties.
In the present work, we resort to the shadow path--integral ground state (SPIGS) \cite{Galli2003,Galli2004} method to compute the imaginary--time intermediate 
scattering function
\begin{equation}
\label{eq:pigs-1}
F({\bf{q}},\tau) = \frac{\langle \Psi_0 | e^{\tau H} \rho_{{\bf q}} e^{-\tau H} \rho_{{\bf -q}}  | \Psi_0 \rangle}{N}
\end{equation}
of the model. In general, $F({\bf{q}},\tau)$ does not have physical meaning per se, but is merely the Laplace 
transform of the dynamical structure factor, 
\begin{equation}
\label{eq:pigs0}
F({\bf{q}},\tau) = \int d\omega \, e^{-\tau \omega} \, S({\bf{q}},\omega) \quad .
\end{equation}
In the case of 2D-SC Bosons, $F({\bf{q}},\tau)$ is related in a surprising and remarkable way to the static density
correlation function of Abrikosov vortexes in Type II superconductor \cite{Nelson1989}.

The problem of inverting equation \eqref{eq:pigs0} to determine $S({\bf{q}},\omega)$ from
$F({\bf{q}},\tau)$ will be discussed
in the forthcoming Subsection \ref{sec:2b}, while the remainder of the present Section is devoted to illustrating
the calculation of \eqref{eq:pigs-1}.

The path--integral ground--state (PIGS) method \cite{Sarsa2000} is based on the notion of imaginary--time
evolution operator $e^{-\tau H}$, where $\tau \geq 0$ is a positive parameter, called imaginary time, 
and $H$ the Hamiltonian of the system.
The PIGS method starts from a variationally optimized trial wavefunction $\Psi_T({\bf{R}})$,
where ${\bf{R}}=({\bf{r}}_1 \dots {\bf{r}}_N)$ denotes the set of coordinates of the particles, and projects
it onto the ground state $\Psi_0({\bf{R}})$ after evolution over a sufficiently long imaginary
time interval $\tau$:
\begin{equation}
\label{eq:pigs1}
\underset{\tau \to \infty}{\lim}
| \Psi_\tau \rangle = 
\underset{\tau \to \infty}{\lim}
\frac{e^{-\tau H} | \Psi_T \rangle}{\| e^{-\tau H} \Psi_T \|} =  
| \Psi_0    \rangle \quad \mbox{if $\langle \Psi_T | \Psi_0 \rangle \neq 0 $} \quad. 
\end{equation}
Projection in imaginary time is accomplished breaking the imaginary--time interval $\tau$
into a large number $M$ of small steps $\delta\tau = \frac{\tau}{M}$, and expressing 
the unknown propagator $G({\bf{R}}',{\bf{R}};\tau) \equiv \langle {\bf{R}}' | e^{-\tau H} | 
{\bf{R}} \rangle$ by means of 
the following convolution formula
\begin{equation}
\label{eq:pigs2}
G({\bf{R}}_M,{\bf{R}}_0;\tau) = \int d{\bf{R}}_{M-1} \dots d{\bf{R}}_1 \prod_{i=0}^{M-1}
G({\bf{R}}_{i+1},{\bf{R}}_i;\delta\tau) \quad .
\end{equation}
The usefulness of this relation stems from the availability of accurate approximations
for the small--time propagator $G({\bf{R}}',{\bf{R}};\delta\tau)$.
In the present work, we have resorted to the well-established pair product approximation
\cite{Pollock1984,Ceperley1995,Yan2015}.
In the light of the representation \eqref{eq:pigs2}, the PIGS estimators of $F({\bf{q}},\tau)$ takes the form
\begin{equation}
\label{eq:pigsfqt}
F({\bf{q}},r\delta\tau) = \frac{ \int d {\bf{X}} \, p({\bf{X}}) \, \rho_{{\bf{q}}}({\bf{R}}_{M+r}) \, 
\rho_{-{\bf{q}}}({\bf{R}}_M) }{ \int d{\bf{X}} \, p({\bf{X}}) }
\quad ,
\end{equation}
${\bf{X}} = ({\bf{R}}_0 \dots {\bf{R}}_{2M})$ being a path in the configuration space of the system, and
\begin{equation}
p({\bf{X}}) = \Psi_T({\bf{R}}_{2M}) \prod_{i=0}^{2M-1}G({\bf{R}}_{i+1},{\bf{R}}_i;\delta\tau) \, \Psi_T({\bf{R}}_0)
\end{equation}
a positive and integrable quantity that can be efficiently sampled using the Metropolis
algorithm \cite{Metropolis1953}.
The PIGS method provides estimates of ground--state properties and imaginary--time correlations functions, which are only
affected by two errors: (i) the use of a finite imaginary time of projection in \eqref{eq:pigs1},
and (ii) the use of a finite time step in \eqref{eq:pigs2}.
The biases introduced by these approximations can be reduced below the statistical uncertainties of the calculation by taking $\tau$ sufficiently large and $\delta\tau$ 
sufficiently small \cite{Rossi2009}.
For the small imaginary--time propagator we have used the pair--product approximation
\cite{Ceperley1995} with imaginary time step $\delta\tau=120 \epsilon^{-1}$
at $\rho^*=0.01$, $\delta\tau=60 \epsilon^{-1}$ at $\rho^*=0.0175$ and 
$\delta\tau=30 \epsilon^{-1}$ at $\rho^*=0.0225$; total imaginary projection time, $\tau$,
ranges from 2400 $\epsilon^{-1}$ to 4800 $\epsilon^{-1}$ depending on the density.
As \eqref{eq:pigsfqt} clearly reveals, $F({\bf{q}},\tau)$ is obtained averaging a product $\rho_{{\bf{q}}}({\bf{R}}_{M+r}) \, 
\rho_{-{\bf{q}}}({\bf{R}}_M)$ of random variables evaluated at distinct imaginary--time instants 
$M$, $M+r$ along the path ${\bf{X}}$. Those instants should be taken in the central part of ${\bf{X}}$, in such a 
way as to ensure complete projection onto the ground state, and to avoid any spurious dependence of
$F({\bf{q}},r\delta\tau)$ on the trial wavefunction.

\subsubsection{Shadow trial wavefunctions and the shadow--PIGS method}

The PIGS method provides asymptotically unbiased estimates of ground--state properties 
and imaginary--time correlation functions for any choice of the trial wavefunction, provided that $\langle \Psi_T | \Psi_0 
\rangle \neq 0$.
The quality of the trial wavefunction, however, has a determining impact on the efficiency
and accuracy of the calculation \cite{Rossi2009}.
Shadow wavefunctions \cite{Vitiello1988,Reatto1988,Vitiello1990,MacFarland1994,Moroni1998} 
(SWFs) take into account interparticle correlations by introducing 
auxiliary variables ${\bf{S}} = ({\bf{s}}_1 \dots {\bf{s}}_N)$, called shadows, 
and coupling them with the real positions ${\bf{R}}$ of the particles through a 
Gaussian kernel:
\begin{equation}
\label{eq:swf1}
\Psi_S({\bf{R}}) = \Psi_{J,r}({\bf{R}}) \int d{\bf{S}} \, e^{- \gamma |{\bf{R}}-{\bf{S}}|^2} \, \Psi_{J,s}({\bf{S}}) \quad,
\end{equation}
where $\Psi_{J,r}$, $\Psi_{J,s}$ are Jastrow wavefunction \cite{MacFarland1994,Vitiello1988}. 
SWFs are particularly flexible, and capable of describing both liquid and solid 
phases depending on the form of the real, shadow and real-shadow correlations \cite{MacFarland1994,Vitiello1988}.
The functional form of such correlation functions is typically determined combining 
physical intuition and mathematical arguments based on the theory of stochastic processes \cite{Caffarel1986,Holzmann2003}.
The parameters in the correlation functions are fixed using on suitable optimization procedures 
\cite{Holzmann2003,Motta2016}.

In the present work, following \cite{Magro1993} and \cite{Rossi2011}, we relied on the 
following real and shadow correlations:
\begin{equation}
\Psi_{J,r}({\bf{R}}) = \prod_{i<j} e^{- u_{J,r}(r_{ij})}
\quad,\quad
u_{J,r}(r) = \left( a+b r^2 \right) \cos\left( \pi d \sqrt{w} r \right) e^{-w r^2}
\end{equation}
and
\begin{equation}
\Psi_{J,s}({\bf{S}}) = \prod_{i<j} e^{- u_{J,s}(s_{ij})}
\quad,\quad
u_{J,s}(s) = a' \, K_0(b' s)
\end{equation}
The shadow--PIGS or SPIGS method \cite{Galli2003,Galli2004} differs from PIGS by projecting in imaginary time a SWF as trial wave function.

\subsection{The GIFT method}
\label{sec:2b}

The output of a SPIGS calculation is a collection 
$\{ F_r \equiv F({\bf{q}},r\delta\tau) \}_{r=0}^{N_\tau-1}$
of estimates of the imaginary--time intermediate scattering function at a finite number 
$N_\tau$ of imaginary time instants. Such estimates are affected by statistical 
uncertainties $\{ \sigma_r \}_{r=0}^{N_\tau-1}$.
The dynamical structure factor should be obtained from those unavoidably limited and noisy estimates
of $F({\bf{q}},\tau)$, by inverting the Laplace transform \eqref{eq:pigs0}.
This is a notoriously ill--posed problem \cite{Tikhonov1977,Tarantola2006,Vitali2010}, in that the Laplace 
transforms of many spectral functions $S({\bf{q}},\omega)$, ranging from featureless to rich--in--structure, 
are compatible with the QMC estimates of $F({\bf{q}},\tau)$.

Several methodologies have been proposed, to determine $S({\bf{q}},\omega)$ \cite{Sandvik1998,Jarrell2007,Mishchenko2012}.
In the present work, we relied on the GIFT method \cite{Vitali2010}, a statistical inversion
method that provides an estimate of $S({\bf{q}},\omega)$ by:
\begin{enumerate}
\item defining a set $\mathcal{S}$ of model spectral functions consistent with any 
prior knowledge about $S({\bf{q}},\omega)$. 
Elements of $\mathcal{S}$ should join ductile form and efficient parametrization. 
In the present work, we have used linear combinations of Dirac delta distributions of the form
\begin{equation}
s(\omega) = S_0 \, \sum_{j=1}^{N_\omega} s_j \, \delta(\omega-\omega_j)
\quad,
\end{equation}
where $S_0=S({\bf{q}})/M$ is the ratio between the static structure factor, $S({\bf{q}})$, 
and a large integer $M$, $N_\omega$ is the number of frequencies, $\omega_j$ are the 
points of a grid with uniform spacing $\delta\omega$, and $s_j$ integer numbers such 
that $\sum_{j=1}^{N_\omega} s_j = M$; in this way the elements of $\mathcal{S}$ 
are normalized to $S({\bf{q}})$ and the large integer $M$ represents the number of
{\em quanta} of spectral weight that the algorithm has to distribute among the
$N_\omega$ frequencies to build a good estimate of the dynamical structure factor.
\item producing a number $N^*$ of equivalent imaginary--time intermediate scattering functions, $F^*_r$, obtained by sampling independent Gaussian distributions, with standard deviations equal to $\sigma_r$, centered on the
original observations $F_r$.
\item defining a fitness function $\Phi: \mathcal{S} \to \mathbb{R}$ that measures 
the compatibility of $s \in \mathcal{S}$ with the QMC data.
In the present work, we used a fitness function enforcing also exact knowledge of the 
first momentum sum rule \cite{Pines1966}:
\begin{equation}
\label{eq:gift_fitness}
\Phi[s] = - \sum_{r=0}^{N_\tau-1} \frac{1}{\sigma_r^2} 
            \left| F^*_r - S_0 \sum_{j=1}^{N_\omega} e^{-r \delta\tau \, \omega_j} s_j \right|^2
          - \gamma \left| \frac{\hbar^2 |{\bf{q}}|^2}{2m} - S_0 \sum_{j=1}^{N_\omega} s_j \omega_j \right|^2
\quad,
\end{equation}
where $\gamma$ is a tunable parameter.
\item devising a genetic algorithm to explore the set $\mathcal{S}$.
\end{enumerate}
The genetic dynamics brought about by such genetic algorithm consists of a succession of generations, during
which a randomly--drawn initial population $\{s_c\}_{c=1}^{N_c}$ of spectral functions is updated applying
suitably--constructed selection, crossover and mutation operations.
Such operations are devised to multiply and propagate across generations spectral functions enjoying higher
compatibility with QMC data, measured by the fitness \eqref{eq:gift_fitness}.

At the end of the calculation, the elements $\{s^*_c\}_{c=1}^{N^*}$ with the highest
fitness in the last generation are averaged, to produce the GIFT estimate
\begin{equation}
S_{GIFT}({\bf{q}},\omega) = \frac{1}{N^*} \sum_{c=1}^{N^*} s^*_c(\omega)
\end{equation}
of the dynamical structure function at a fixed wavevector ${\bf{q}}$.
The GIFT method has been successfully applied to the study of $S({\bf{q}},\omega)$ 
in 1D, 2D and 3D systems of He atoms \cite{Bertaina2016,Vitali2010,Arrigoni2013,Nava2013} and 3D 
systems of hard spheres \cite{Rota2013}.
In this work, typically we have used: ${N_c}=2\times 10^4$, $M=5\times 10^3$, 
$N_\omega=600$, $N^*=500$ and 
$\delta\omega=2.8 \times 10^{-5} \epsilon$ at $\rho*=0.01$,
$\delta\omega=5.6 \times 10^{-5} \epsilon$ at $\rho*=0.0175$,
$\delta\omega=1.12 \times 10^{-4} \epsilon$ at $\rho*=0.0225$.
The interested reader is deferred to \cite{Vitali2010,Rota2013} for further technical details.

\section{Static properties}
\label{sec:3}

\begin{figure}[t]
\begin{center}
\includegraphics*[width=0.4\textwidth]{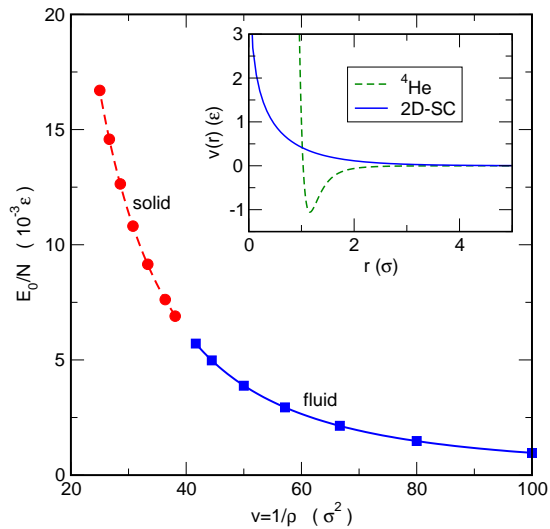}
\caption{(color online) Energy per particle of 2D-SC Bosons
\cite{Zavattari2008}.
Red (blue) symbols correspond to the fluid (solid) phase. Where not shown, statistical uncertainties are below the symbol size.
In the inset, a comparison between an accurate interaction potential among He atoms and in the
2D-SC system is shown in reduced units ($\sigma_{He}=2.556$~\AA, $\epsilon_{He}=10.22$ K).}
\label{fig:eqstate}
\end{center}
\end{figure}

\begin{figure}[t]
\begin{center}
\includegraphics*[width=0.4\textwidth]{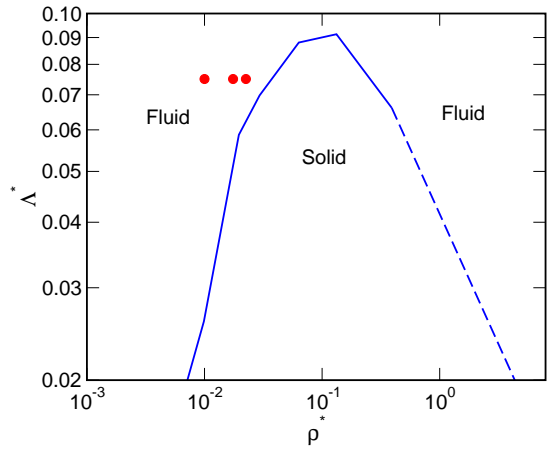}
\caption{(color online) Phase diagram of 2D-SC Bosons, from \cite{Magro1993}. The red points correspond to the three systems, studied in the present work, where we have computed $F({\bf{q}},\tau)$, $S({\bf{q}},\omega)$ and one--body density--matrix.}
\label{fig:punti_ES}
\end{center}
\end{figure}

As mentioned in the introduction, the $T=0$ K phase diagram of the 2D-SC Bosons in the ($\Lambda^*$,$\rho^*$)
plane is characterized by a dome below which the stable phase is a solid, and above which it is a fluid.
At fixed $\Lambda^*$, and for $\rho^*$ in the vicinity of the transition, the energies of the fluid and of the 
solid phase are very close and with a similar curvature. This fact is exemplified in Figure~\ref{fig:eqstate}, where we show the ground--state
energy per particle for $80$ Bosons at $\Lambda^* = 0.075$, and makes rather difficult to accurately locate
the fluid-solid transition with the Maxwell construction.
As a consequence the phase diagram, sketched in Figure~\ref{fig:punti_ES}, is known only with limited accuracy.

In the present work, we have simulated systems of $N=80$ 2D-SC Bosons at various reduced densities 
with the DeBoer parameter fixed to $\Lambda^* = 0.075$. As illustrated in Figure \ref{fig:punti_ES},
the investigation of diagonal, off--diagonal and dynamical properties has been performed at three different reduced 
densities $\rho^* = 0.01$, $0.0175$, $0.0225$.
As the density increases, the crystallization of the system is approached.
For each of the considered densities, we have computed the static structure factor
\begin{equation}
S({\bf{q}}) = \int d\omega \, S({\bf{q}},\omega) = F({\bf{q}},0) \quad ,
\end{equation}
the radial distribution function $g(r)$, related to the static structure factor by
\begin{equation}
S({\bf{q}}) = 1 + \rho \int d{\bf{r}} \, e^{- i {\bf{q}} \cdot {\bf{r}}} \, g({\bf{r}}) \quad,
\end{equation}
and the off--diagonal one--body density matrix
\begin{equation}
\rho_1({\bf{r}},{\bf{r}}') = N \int d{\bf{r}}_2 \dots {\bf{r}}_N \Psi^*({\bf{r}}  \, {\bf{r}}_2 \dots {\bf{r}}_N)
                                                             \Psi  ({\bf{r}}' \, {\bf{r}}_2 \dots {\bf{r}}_N)
\quad ,
\end{equation}
as well as the imaginary--time intermediate scattering function \eqref{eq:pigs-1} and the dynamical structure factor \eqref{eq:pigs0}, 
which is the central observable of the present work.
In the fluid phase $\rho_1({\bf{r}},{\bf{r}}')$ is function of the distance $|{\bf{r}}-{\bf{r}}'|$ and its Fourier 
transform represents the momentum distribution. 
When $\rho_1(|{\bf{r}}-{\bf{r}}'|)$ does not vanish at large distance, there is a Bose-Einstein condensate (BEC) 
and the quantity $n_0 \equiv \lim_{r \to \infty} \rho_1(r)/\rho$ represents the fraction of particles in the BEC.

\begin{figure}
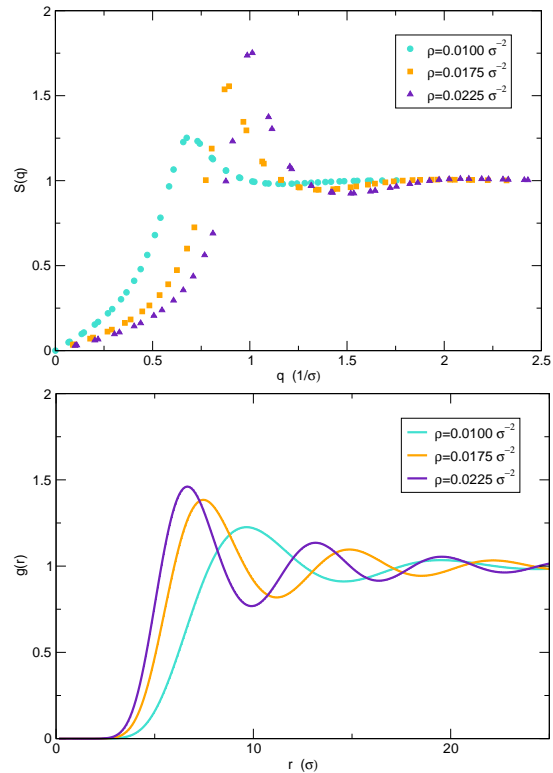

\begin{center}
  \begin{minipage}{0.4\textwidth}
    \includegraphics*[width=\textwidth]{essediq.eps}
  \end{minipage}
\\
  \begin{minipage}{0.4\textwidth}
    \includegraphics*[width=\textwidth]{gdir.eps}
  \end{minipage}
\caption{(color online) Static structure factor (upper panel) and radial distribution function (lower panel) for systems of $N=80$
2D-SC Bosons at $\Lambda^* = 0.075$ and $\rho^* = 0.01$, $0.0175$, $0.0225$ (light blue, orange, purple respectively). Where not shown, statistical uncertainties are below the symbol or line size.}
\label{fig:essediq}
\end{center}
\end{figure}

For $\Lambda^* = 0.075$ the densities of the present study are in the normal region, 
in that the short--range order increases with the density. 
This is shown in Fig.\ref{fig:essediq}: the oscillations of $g(r)$ and the peak of $S(q)$ increase and sharpen as the density increases.
This phenomenon is reminiscent of hard--core systems like ${}^4$He.

At the same time there is a displacement in the position of the peaks: the position $r_1$ of the first peak of $g(r)$ decreases with
increasing density, and the position $q_{M,S}$ of the main peak of $S(q)$ increases for increasing density, as detailed in
Table \ref{tab:1}.
Here we can notice a quantitative difference with respect to $^4$He. For 2D-SC the position of the peak of $S(q)$ increases 
by 45\% as the density changes from 0.01 to 0.0225 and this displacement is very close to the square root of the ratio of those two 
densities.
This means that the variation of the short--range order of the 2D-SC fluid corresponds closely to what one would expect 
from a simple compression of the positions of the particles. In the case of ${}^4$He, such displacements are much smaller. For instance, 
in 2D ${}^4$He the freezing density is about 50\% larger of the equilibrium density \cite{Arrigoni2013}, so that a simple compression 
gives a $\sqrt{1.5}=1.22$ displacement of the peaks of $g(r)$ and $S(q)$.
The actual displacements of these peaks for $^4$He in 2D are instead only 8\%, much smaller of the value resulting from compression.
This difference between the two systems can be attributed to the different stiffness of the short--range repulsion of the two systems (see inset in Figure \ref{fig:eqstate}). 
As discussed in the forthcoming Section, this different density dependence of the short--range order has a dramatic effect on the density 
dependence of the excitation spectrum of soft--core particles, compared to hard--core particles.

\begin{figure}
\begin{center}
  \begin{minipage}{0.39\textwidth}
    \includegraphics[width=\textwidth]{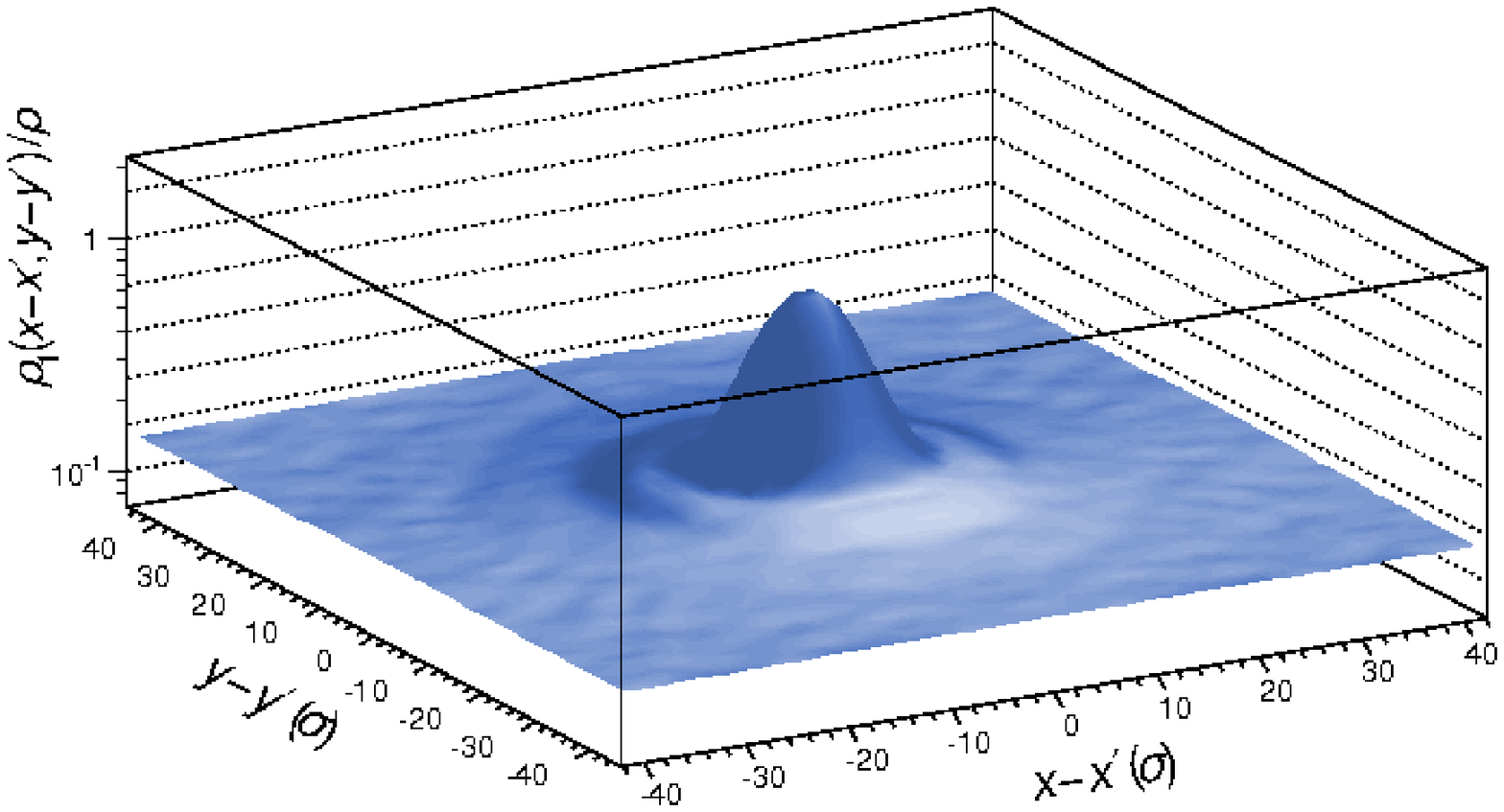}
  \end{minipage}
  \begin{minipage}{0.39\textwidth}
    \includegraphics[width=\textwidth]{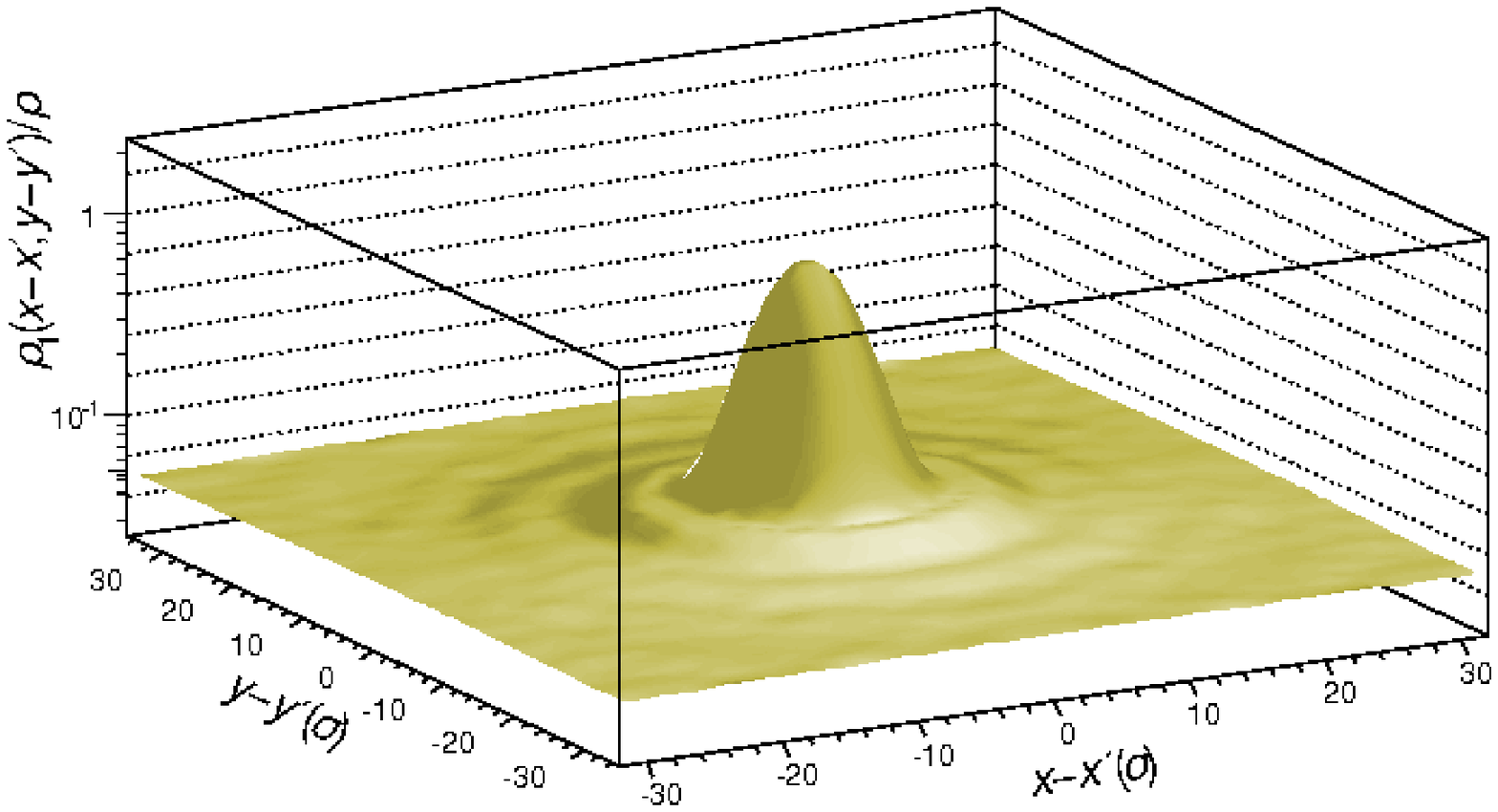}
  \end{minipage}
\\
  \begin{minipage}{0.39\textwidth}
    \includegraphics[width=\textwidth]{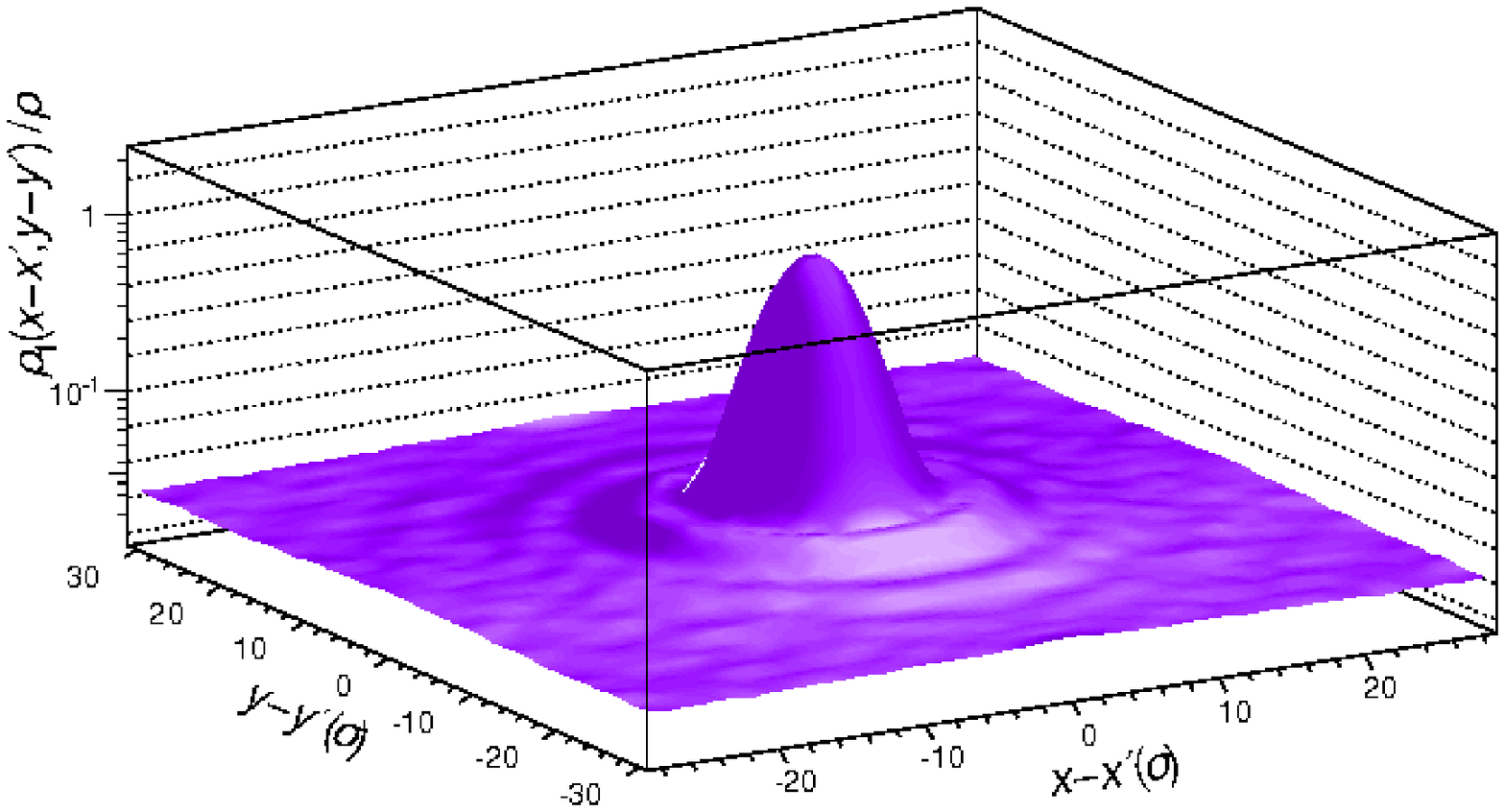}
  \end{minipage}
    \begin{minipage}{0.39\textwidth}
    \includegraphics*[width=\textwidth]{odlro_rad_new.eps}
  \end{minipage}
\caption{(color online) One--body density matrix for systems of $N=80$
2D-SC Bosons at $\Lambda^* = 0.075$ and $\rho^* = 0.01$ (left, top), $0.0175$
(right, top), $0.0225$ (left, bottom).
Radial plot of the one--body density matrix for the studied densities (right, bottom).}
\label{fig:obdm}
\end{center}
\end{figure}

The results for the off--diagonal one--body density matrix are shown in Fig.\ref{fig:obdm}.
The prominent feature is the plateau at large distance, reflecting the emergence of BEC. 
Therefore the present system is superfluid. As expected $n_0$ decreases as the system becomes more strongly coupled, due to the increase
in the density. 
In fact the condensate is 13.3\% at $\rho=0.01$, 4.7\% at $\rho=0.0175$ and 2.5\% at $\rho=0.0225$. We remind that 100\% of the particles are in the condensate in 
the limit of vanishing density. The trend of the condensate with density is similar to that observed in ${}^4$He, both in 2D and 3D.
For instance, the BEC fraction of $^4$He in 2D changes from 21\% at equilibrium density to 3\% at freezing. 

One might ask whether there is a general trend of the condensate fraction, as the system becomes increasingly coupled.
To this purpose, let us recall that a measure of the degree of local order is provided by the height $S_M$ of the first peak of $S(q)$.
In Figure~\ref{fig:5}, the condensate fraction $n_0$ is shown as function of $S_M$ for 2D-SC and for $^4$He in 2D. 
It clearly emerges a surprising similarity in the behavior of $n_0$ for these two very different systems.

Once divided by the reduced density, $\rho_1(r)$ starts from unity at $r=0$ and decays as a quadratic function of $r$ with curvature 
proportional to the kinetic energy. At intermediate distances, before reaching the long--distance BEC plateau, $\rho_1(r)$ has some weak 
damped oscillations, that are slightly stronger at larger density.
This is similar to the behavior of $\rho_1(r)$ in $^4$He.

\section{Dynamical properties}
\label{sec:4}

\begin{figure}
\begin{center}
  \begin{minipage}{0.55\textwidth}
    \includegraphics*[width=\textwidth]{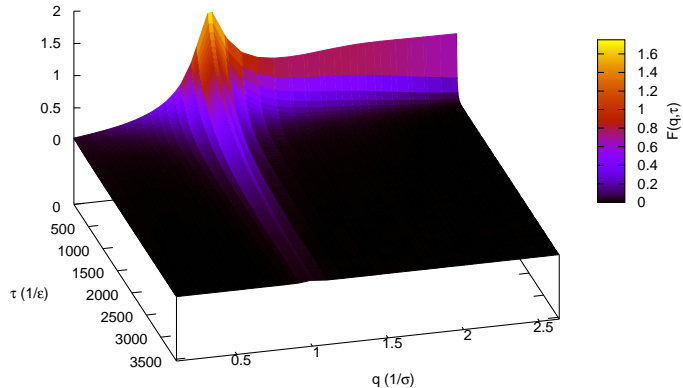}
  \end{minipage}
\caption{(color online) $F({\bf{q}},\tau)$ 
for $N=80$ 2D-SC Bosons at $\Lambda^* = 0.075$ and $\rho^* = 0.0225$}
\label{fig:fq3d}
\end{center}
\end{figure}
\begin{figure}
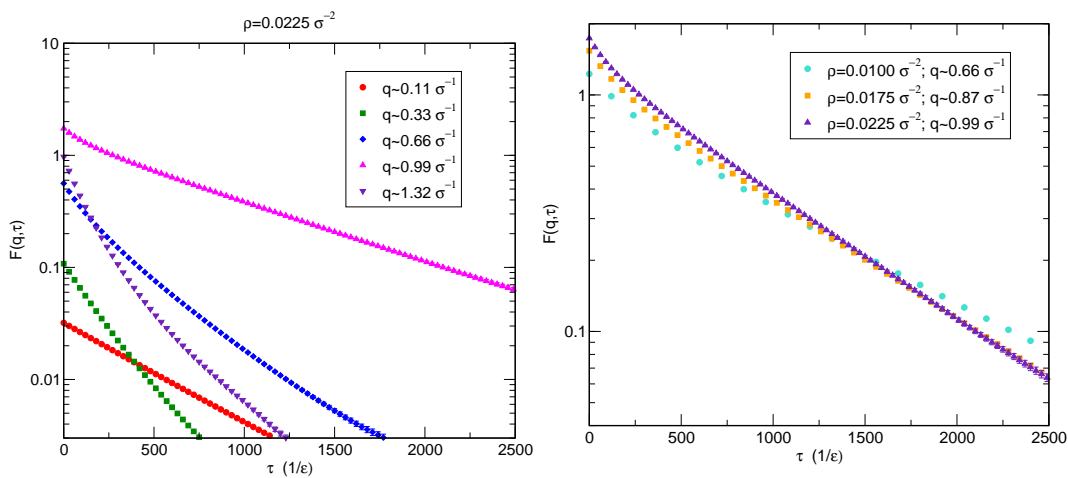

\begin{center}
  \begin{minipage}{0.39\textwidth}
    \includegraphics*[width=\textwidth]{effetauqu_new.eps}
  \end{minipage}
  \begin{minipage}{0.39\textwidth}
    \includegraphics*[width=\textwidth]{effetau.eps}
  \end{minipage}
\caption{(color online) Logarithmic plot of $F(q,\tau)$ for selected values of $q$ at 
one of the studied densities (left) and at all the studied densities at the roton
wave--vector (right).}
\label{fig:fdiq}
\end{center}
\end{figure}

The outputs of our quantum simulations are maps of the imaginary--time 
intermediate scattering function $F(q,\tau)$ like that illustrated 
in Figure~\ref{fig:fq3d}.
The computed $F(q,\tau)$ is an exact statistical representation of this quantity. 
Unfortunately the Laplace transform (6) that relates $F(q,\tau)$ to the dynamical structure factor $S(q, \omega)$ is an ill--posed mathematical problem for which powerful but 
approximate methods have been devised as mentioned in Section \ref{sec:2}.
If $S(q, \omega)$ has a sharp peak corresponding to well--defined excitations like the 
phonon--maxon--roton in superfluid $^4$He, it is known \cite{Vitali2010} that the GIFT inversion method we are using gives an accurate characterization of such excitations. 
Results suggest this is true also for the 2D-SC system under investigation.
Logarithmic plots of $F(q,\tau)$ as function of $\tau$ for selected values of $q$ at one of the studied densities are displayed in Figure~\ref{fig:fdiq}. 
At small $q$, $\ln F(q,\tau)$ is essentially a straight line in $\tau$ and an exponential decay of $F(q,\tau)$ corresponds to the presence of a sharp peak in $S(q,\omega)$ with 
the $\omega$ of the peak representing the energy $\epsilon(q)$ of a well--defined excitation of the system. Also for $q$ values close to the main peak of $S(q)$ the function
$\ln F(q,\tau)$ displays a linear behavior in $\tau$ over an extended range of $\tau$ values, as it can be seen in Figure~\ref{fig:fdiq}, and only at small $\tau$ some
deviation from the linear behavior can be noticed.

\begin{figure}
\begin{center}
  \begin{minipage}{0.45\textwidth}
    \includegraphics[width=\textwidth]{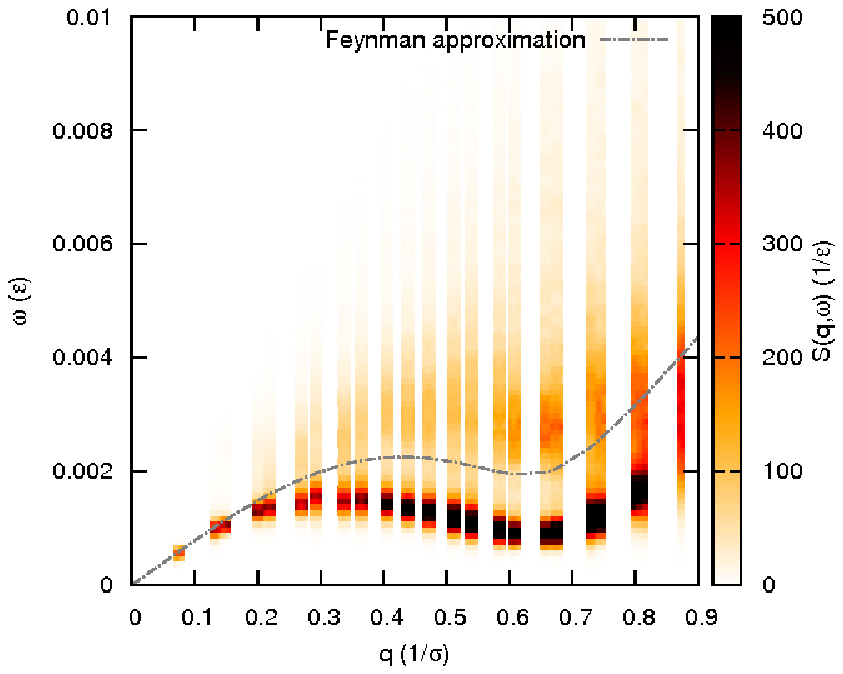}
  \end{minipage}
  \\
  \vskip -9mm
  \begin{minipage}{0.45\textwidth}
    \includegraphics[width=\textwidth]{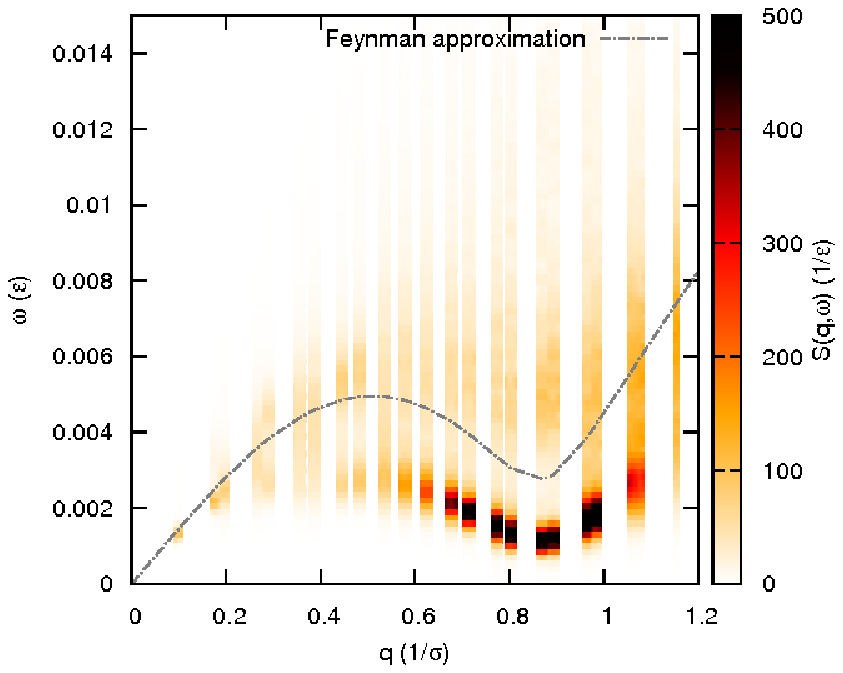}
  \end{minipage}
  \\
  \vskip -9mm
  \begin{minipage}{0.45\textwidth}
    \includegraphics[width=\textwidth]{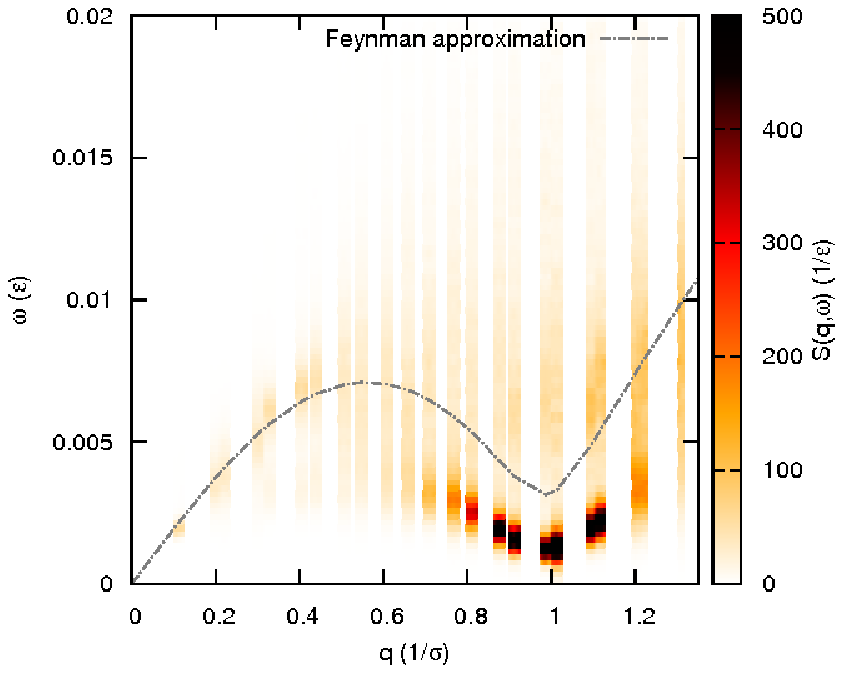}
  \end{minipage}
\caption{$S({\bf{q}},\omega)$ for $N=80$
2D-SC Bosons at $\Lambda^* = 0.075$ and $\rho^* = 0.01$ (upper panel), $\rho^* = 0.0175$ (central panel), $\rho^* = 0.0225$ (lower panel). (dot-dashed lines)
excitation energies computed with the Feynman approximation (\ref{eq:20}).}
\label{fig:dyn}
\end{center}
\end{figure}

Therefore also for $q$ in the region of the main peak of $S(q)$ well--defined excitations 
are present. For other values of $q$, $\ln F(q,\tau)$ shows substantial deviation from linearity. The GIFT method allows to estimate $S(q,\omega)$ also in such situations,
and the results are shown in Figure~\ref{fig:dyn} at the three studied densities. 

\begin{figure}
\begin{center}
  \begin{minipage}{0.45\textwidth}
    \includegraphics[width=\textwidth]{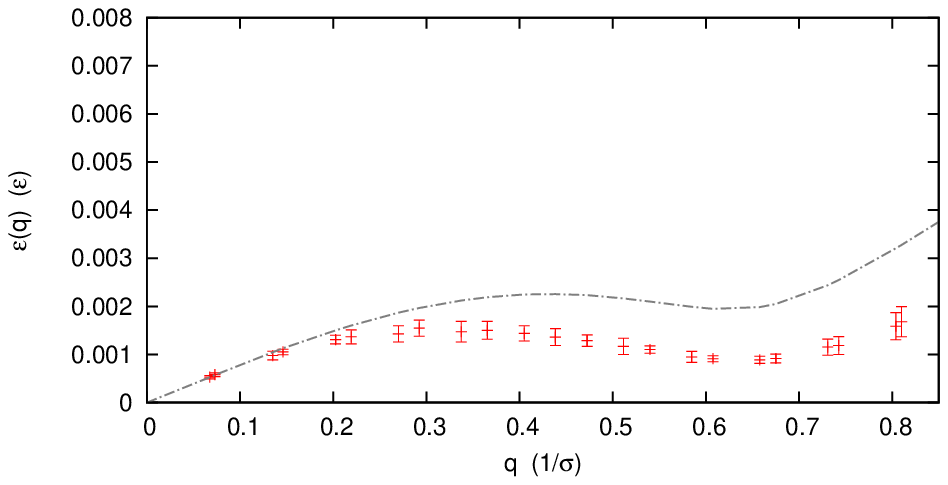}
  \end{minipage}
  \\
  \begin{minipage}{0.45\textwidth}
    \includegraphics[width=\textwidth]{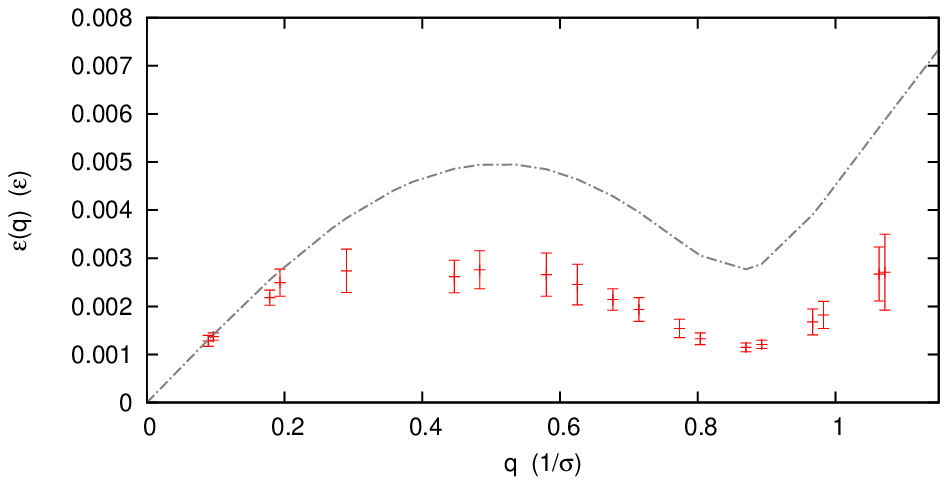}
  \end{minipage}
  \\
  \begin{minipage}{0.45\textwidth}
    \includegraphics[width=\textwidth]{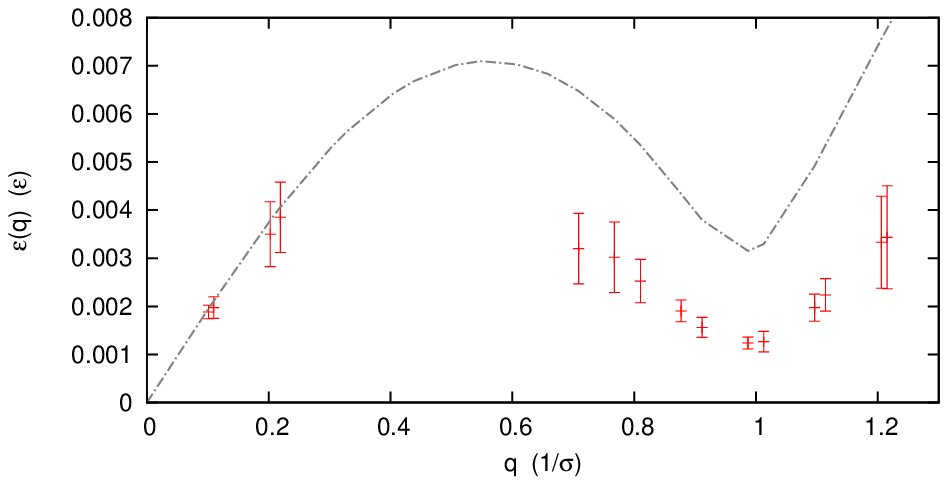}
  \end{minipage}
\caption{(color online) $\epsilon({\bf{q}})$ for $N=80$ 2D-SC Bosons at 
$\Lambda^* = 0.075$ and $\rho^* = 0.01$ (upper panel), $\rho^* = 0.0175$ (central panel), 
$\rho^* = 0.0225$ (lower panel); Feynman's approximations (dot-dashed lines) to $\epsilon({\bf{q}})$ are also shown.}
\label{fig:9}
\end{center}
\end{figure}

At small $q$, $S(q,\omega)$ consists essentially of a single sharp peak in $\omega$. Sharp peaks are present also for $q$ values in the region 
of the maximum of $S(q)$ and the energy $\epsilon(q)$ has a minimum $\Delta_R=\epsilon(q_R)$ 
at a finite wave vector $q_R$ as shown in Figure~\ref{fig:9}, where 
$\epsilon(q)$ is plotted as function of $q$. 
By analogy with $^4$He we call {\em rotons} such excitations. At small $q$, $\epsilon(q)$ 
is linear in $q$, revealing phonon excitations.
The error bars in Figure~\ref{fig:9} represent the full width of the excitation peaks in the reconstructed $S(q,\omega)$. 
Such finite width has twofold origin: even in the case of a Dirac delta spectral distribution, the reconstructed $S(q,\omega)$ shows a peak of 
finite width \cite{Vitali2010}, in which case the width is a measure of the statistical uncertainty on the excitation energy.
On the other hand, an excitation can have a finite lifetime. In this case the width of the reconstructed peak of $S(q,\omega)$ depends on both 
the finite lifetime and the statistical uncertainty on the excitation energy.
To the best of our knowledge, it is not possible to discern the two contributions.
In the case of rotons, many--body theories \cite{Fetter1971,Glyde1994} indicate that these excitations at $T=0$ K have infinite lifetime, so the width shown in the
Figure should represent the statistical uncertainty in the computation of the roton energy.

At $q_R$, in addition to the sharp peak, $S(q,\omega)$ has broad contributions at high $\omega$.
Also $^4$He has broad contributions at large $\omega$ \cite{Arrigoni2013}, that can be interpreted as due to multi--phonon processes \cite{Fetter1971,Glyde1994}.
We recall that the integral of $S(q,\omega)$ over all $\omega$ is equal to $S(q)$. Therefore the ratio $Z(q)/S(q)$, where $Z(q)$ is the integral of 
$S(q,\omega)$ over the $\omega$ region of the sharp peak, represents the weight of the elementary excitation contribution to $S(q,\omega)$. 
The strength $Z(q)$ of the elementary excitation is computed by numeric integration of $S(q,\omega)$
around the peak.
All the data reported in Table \ref{tab:1} are relative to situations where the elementary excitation 
peak is well separated from the high--energy multiphonon structure, ensuring a clear identification of 
the integration domain.
At small $q$, $Z(q)/S(q)$ is close to unity whereas for the roton $Z(q)/S(q)$ is about 0.6-0.7, and it is an increasing function of density.
This is similar to $^4$He and some results for $^4$He in 2D and for the 2D-SC system are reported in Table \ref{tab:1}.

\begin{table}
\begin{center}
\begin{tabular}{cccccccccc}
\hline\hline
$\rho^*$ & $E/N$      & $T/N$      & $q_{M,S}$ & $q_{M,Z}$ & $q_R$ & $S_M$ & $Z_M$ & $Z_M/S_M$ \\
\hline
0.0100   & 0.96(2) & 0.58(2) & 0.69        & 0.69        & 0.64  & 1.27      & 0.79      & 0.622 \\
0.0175   & 2.94(3) & 1.40(3) & 0.89        & 0.88        & 0.86  & 1.56      & 1.12      & 0.718 \\
0.0225   & 4.99(7) & 2.06(7) & 1.00        & 1.00        & 0.98  & 1.77      & 1.34      & 0.757 \\
\hline\hline
\end{tabular}
\end{center}
$ $

$ $
\begin{center}
\begin{tabular}{cccccccccc}
\hline\hline
$\rho$  & $E/N$     & $T/N$    & $q_{M,S}$ & $q_{M,Z}$ & $q_R$ & $S_M$ & $Z_M$ & $Z_M/S_M$ \\
\hline
0.04315 & –0.859(2) & 3.892(4) & 1.63        & 1.59        & 1.37  & 1.24      & 0.82      & 0.661  \\
0.0536  & –0.704(2) & 5.744(4) & 1.67        & 1.65        & 1.55  & 1.41      & 1.01      & 0.716  \\
0.0658  &  0.062(3) & 8.656(6) & 1.76        & 1.75        & 1.73  & 1.78      & 1.33      & 0.747  \\
\hline\hline
\end{tabular}
\end{center}
\caption{Comparison of the condensate fraction, total and kinetic energy and the ratio $Z(q)/S(q)$ for 2D SC Bosons and 2D $^4$He.
$q_{M,S}$ and $q_{M,Z}$ are the wavevectors at which $S(q)$ and $Z(q)$ attain the maximum values
$S_M$ and $Z_M$ respectively, and $q_R$ is the roton wavevector.
Part of the $^4$He data are taken from \cite{Arrigoni2013}.
Wavevectors (densities,energies) are measured in units of $\sigma^{-1}$ ($\sigma^{-2}$, $10^{-3} \, \epsilon$) for 2D SC Bosons and 
$\AA^{-1}$ ($\AA^{-2}$, $K$) for 2D $^4$He respectively.
} \label{tab:1}
\end{table}

At wavevectors $q$ between the phonon and the roton region, $S(q,\omega)$ has a maximum at $q$ about $q_R/2$, in suggestive analogy with the 
so--called maxon region in $^4$He.
At the lowest density $\rho^* = 0.01$ the peak of $S(q,\omega)$ is rather sharp, so we have a well--defined maxon excitation. At the two higher densities, 
$\rho^* = 0.0175$ and $0.0225$, only a very broad peak is present at such values of $q$.
Therefore the 2D-SC system at high density has well--defined collective excitations only in a restricted 
region of $q$, phonons at small $q$ and rotons in a finite region centered at $q_R$.
This behavior has a simple explanation: when the maxon energy is larger than twice the
roton energy $\Delta_R$, the maxon can decay in two rotons, and thus acquires a finite 
lifetime even at $T=0$ K \cite{Glyde1994}. When there is overdamping $S(q,\omega)$ has only a very broad peak as the remnant of the excitation. At $\rho^*=0.01$ 
the maxon energy is lower than $2\Delta_R$, so that the maxon is stable as shown by the sharp peak in Figure~\ref{fig:9}. 
The converse is true at the two other densities, where the maxon feature is well above $2\Delta_R$ and the corresponding excitation is no longer
stable.
This is similar to $^4$He where, close to the freezing density, the maxon gets a finite lifetime 
even at the lowest temperature of experiment \cite{Glyde1994}.
In 2D-SC system the situation is more extreme: due to the broad density range in which the fluid state is stable, the remnant of the maxon excitation 
is at an energy many time larger of the roton energy.

The excitation energy $\epsilon(q)$ of the 2D-SC system at the three densities of our
computation is plotted as function of $q$ in Figure~\ref{fig:9}.
Here a striking difference with respect to $^4$He is observed in the density dependence of the roton energy $\Delta_R$.
In 3D $^4$He it is known experimentally that $\Delta_R$ is a decreasing function of $\rho$,
and the same behavior is known to be true from quantum simulations of $^4$He in 2D \cite{Arrigoni2013}.
Such behavior is so characteristic that often the roton excitation is considered as the soft mode of the solidification phase transition 
\cite{Enz,Nozieres}, the vanishing of $\Delta_R$ being preempted by the first--order nature of the 
fluid--solid transition. Our results show that the roton energy of the 2D-SC system 
is an {\em increasing} function of density as solidification is approached. 
This jeopardizes the notion of the roton as the soft-mode of the fluid-solid transition 
as a generic property of this quantum phase transition, and represents one of the main results of our computations.
It is possible to understand the origin of the different density dependence of the 
roton energy in our soft--core system and in $^4$He. 
Consider the Feynman spectrum of excitations in a Bose fluid,
\begin{equation}\label{eq:20}
\epsilon_F(q)={\hbar^2 q^2 \over 2 m S(q)} \quad .
\end{equation}
We remind the first few sum rules of $S(q,\omega)$ at $T=0$ K:
\begin{equation}\label{eq:21}
m_0=S(q)\, , \quad m_1={\hbar^2 q^2 \over 2 m}\, , \quad m_{-1}=-{\chi(q) \over 2 \rho}
\end{equation}
The Feynman spectrum (\ref{eq:20}) can be obtained from the first two sum rules 
in (\ref{eq:21}) under the assumption that $S(q,\omega)$ is equal to a Dirac delta 
function, i.e. $S(q,\omega)=Z(q)\delta(\omega-\epsilon(q))$. 
Naturally, this is an approximation and the actual excitation spectrum in the roton 
region is well below the prediction of (\ref{eq:20}) which is also plotted in Figure~\ref{fig:9}. 
The depression of $\Delta_R$ below $\epsilon_F(q_R)$ is a measure of many--body effects 
going under the name of backflow \cite{Feynman}.
Let us now consider the density dependence of $\epsilon_F(q_R)$: $q_R$ is an increasing function of density,
and the maximum of $S(q)$ is an increasing function of the density.
Therefore both numerator and denominator of the Feynman spectrum (\ref{eq:20}) are increasing functions of density, 
so that the density dependence of the roton energy depends on the relative importance of those effects.
In $^4$He, $q_R$ depends weakly on $\rho$ and the increase of $S_M$ is considerable, so that the 
roton energy is a decreasing function of $\rho$. In 2D-SC, $q_M$ has a stronger density 
dependence compared to $^4$He, due to the softness of the repulsion at short range:
the density dependence of the numerator in (\ref{eq:20}) is the dominant effect, whence the roton energy is 
an increasing function of $\rho^*$.
To this effect, one has to add the 
backflow contribution to the roton energy which is an increasing function of $\rho$.
The dependence of the roton energy on the density is thus determined by two competing effects, on which
the details of the interatomic potential have a strong and non--trivial impact.
As a consequence, there is no general dependence of the roton energy on the density.

\begin{figure}[t]
\begin{center}
\includegraphics*[width=0.4\textwidth]{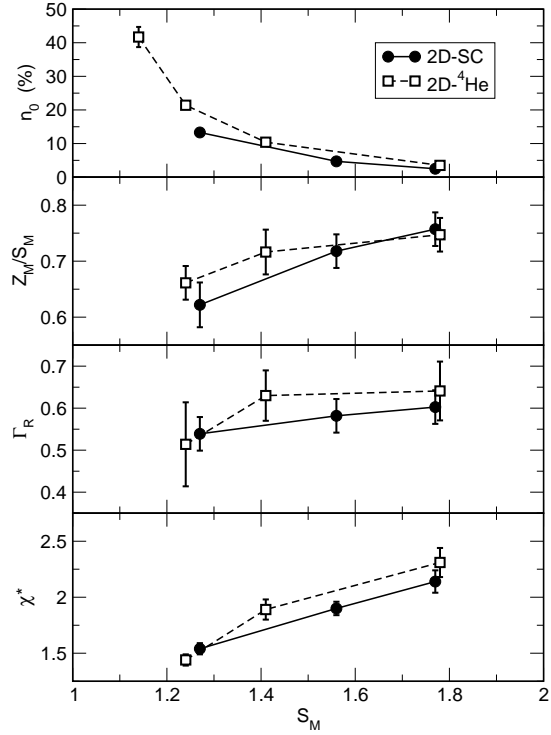}
\caption{Condensate fraction $n_0$, ratio $Z_M/S_M$ between the maximum of $Z(q)$ 
and $S(q)$, backflow effect $\Gamma_R$ \eqref{eq:22}, and ratio $\chi^*$ between the 
maximum of $\chi(q)$ and the maximum of $\chi_F(q)$, as a function of the maximum of $S(q)$, $S_M$ for the 2D-SC and the 2D $^4$He systems. 
Statistical uncertainties on the values of $S_M$ are below the symbols size.
}
\label{fig:5}
\end{center}
\end{figure}

\begin{figure}[t]
\begin{center}
\includegraphics*[width=0.4\textwidth]{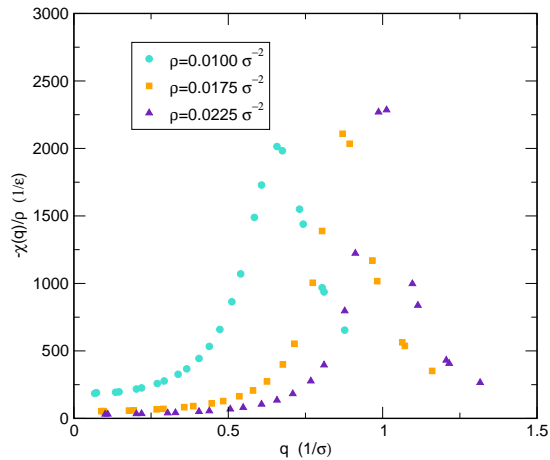}
\caption{(color online) Density response function extracted from the dynamical
structure factor at different densities and $\Lambda^* = 0.075$}
\label{fig:chiq}
\end{center}
\end{figure}

One question one might ask is if the backflow effect on the roton energy depends on 
the nature of the interatomic potential or not. A measure of this backflow effect is 
\begin{equation}\label{eq:22}
\Gamma_R=(\epsilon_F(q_R)-\Delta_R)/\epsilon_F(q_R)
\end{equation}
and this quantity is plotted in Figure~\ref{fig:5} as function of the height of the maximum of $S(q)$ for $^4$He 
in 2D and for 2D-SC. One can notice that the backflow effect on the roton energy increases with density in a similar way 
in these two systems.
Another quantity of interest in a many-body system is the density--density response function $\chi(q)$ to a static external
potential of wave--vector $q$.
We obtain $\chi(q)$ from the third sum rule in (\ref{eq:21}) using the reconstructed $S(q,\omega)$. In $^4$He, $\chi(q)$ 
is characterized by a large peak at a wave vector close to that of the maximum of $S(q)$.
As shown in Figure~\ref{fig:chiq}, the density--density response function $\chi(q)$ of 2D-SC has a similar behavior. A simple 
estimate of $\chi(q)$ is obtained from the Feynman approximation \cite{Feenberg}:
\begin{equation}\label{eq:23}
m_{-1} = -\frac{\chi_{F}(q)}{2\rho}
= \frac{S(q)}{\epsilon_F(q)} = 2m \left[ \frac{S(q)}{\hbar q} \right]^2
\end{equation}
A measure of the many-body effects on $\chi(q)$ is the 
ratio, $\chi^{\star}$, of the computed maximum $\chi_M$ of $\chi(q)$ and of the maximum of $\chi_F(q)$. 
The values of $\chi^{\star}$ for the 2D-SC system and for $^4$He in 2D are plotted as function 
of the maximum of $S(q)$ in Figure~\ref{fig:5}.
Also for this quantity there is similarity among these two systems.

\section{Conclusions}
\label{sec:5}

We have presented a study of static and dynamical properties at $T=0$ K of Bosons 
interacting with the 2D screened Coulomb potential along a constant value 
$\Lambda^*=0.075$ of the DeBoer parameter as function of density as the fluid 
approaches solidification. For the range of density of our study solidification takes 
place upon increasing density like in $^4$He. 
One purpose of our study was to identify similarities and differences between a hard core system 
like $^4$He and  a soft core system like 2D-SC. 
We find that similarities between these two systems in 2D are indeed present but 
our results highlight also substantial differences. Both systems have BEC and are superfluid and the 
short range spatial order  is an increasing function of density. 
However in 2D-SC the range of this order as measured, for instance, by the q-vector of the first peak of S(q) has a much stronger density dependence compared to that of $^4$He. 
On the basis of our powerful GIFT inversion method of the imaginary-time intermediate 
scattering function we are able to get a quantitative evaluation of the dynamical 
structure factor and of the spectrum $\epsilon(q)$ of the excitations of the system. 
$\epsilon(q)$ of 2D-SC has the typical phonon-maxon-roton structure as was suggested 
earlier \cite{Nordborg1997,Nordborg1998}. Phonons and rotons are sharp excitations at all densities of our computation. 
Unlike in the case of $^4$He, maxons are well defined excitations only for density rather
far from solidification.

The most striking difference from 
$^4$He is found in the density dependence of the roton energy $\Delta_R$ . 
In $^4$He $\Delta_R$ is strongly depressed by increasing density whereas in 2D-SC 
$\Delta_R$ increases as $\rho$ approaches solidification. 
Therefore the view \cite{Enz,Nozieres} of the roton as the soft mode of the fluid-solid 
transition cannot be taken as a general description of the solidification transition 
of Bosons. Notwithstanding  such differences, we find that it is possible to put 
into quantitative correspondence a number of properties of these two systems if such quantities are expressed in terms of the height $S_M$ of the first maximum of $S(q)$. 
We have verified this for the condensate, for the strength of the roton peak in 
$S(q,\omega)$ and for the backflow contribution to the roton energy and to the 
maximum of the density--density response function. This last property is quite 
remarkable because one might have expected that backflow should be weaker in a 
soft core system like 2D-SC.

Our findings open a number of questions. At $\Lambda^*=0.075$ by increasing 
further the density one first finds the solid phase but this solid melts at 
still higher $\rho$ (see fig. 2). One question is what happens to the excitation 
spectrum at this inverted solidification transition, is there a kind of symmetry 
compared with the normal solidification at lower density or not? Another question 
is how this high density fluid evolves toward the 2D unscreened Coulomb system 
in which \cite{Nordborg1997,Nordborg1998} phonons are replaced by plasmons and BEC is suppressed to zero. 
Above a critical value of $\Lambda^*$ solidification disappears and the system 
is fluid at all densities. It will be interesting to investigate how the dynamical 
properties evolve, presumably in a non-monotonic way as one moves from low density to very large $\rho$.
We have shown that some properties of 2D-SC and $^4$He can be put into quantitative correspondence for states characterized by the same amount of short range order. 
It should be interesting to verify which is the degree of validity of such 
quasi-universality by changing the value of $\Lambda^*$ of 2D-SC and by studying 
other strongly interacting Bosonic fluids in 2D, as well as an investigation of the 
possible validity of this law also for Bosonic fluids in 3D.

\begin{acknowledgements}
It is a pleasure to dedicate this contribution to Flavio Toigo, honoring a scientist who has given important contributions in many area of condensed matter and statistical physics.
We acknowledge the CINECA and the Regione Lombardia award, under the LISA initiative, for the availability of high performance computing resources and support.
One of us (L.R.) wants to thank Dipartimento di Fisica, Universit\`a degli Studi di Milano, for some support to his research activity. 
One of us (M.M.) acknowledges support from Dipartimento di Fisica, Universit\`a degli Studi di Milano, the Simons Foundation and NSF (Grant no. DMR-1409510).
\end{acknowledgements}

\end{document}